\documentclass[aps,prl,superscriptaddress,twocolumn]{revtex4-1}

\usepackage{mathrsfs}
\usepackage{amsfonts}
\usepackage{amssymb}
\usepackage{amsmath}
\usepackage{graphicx}
\usepackage[usenames,dvipsnames]{color}
\usepackage[colorlinks=true,citecolor=blue,linkcolor=magenta]{hyperref}
\usepackage{lmodern}
\usepackage[euler]{textgreek}

\newcommand{\figpath}{.}

\newcommand{\be}{\begin{equation}}
\newcommand{\ee}{\end{equation}}
\newcommand{\pd}[2]{ \frac{\partial{#1}}{\partial{#2}}}

\newcommand{\kB}{k_\mathrm{B}}

\begin{document}

\title{Detecting continuous spontaneous localisation \\ with charged bodies in a Paul trap}

\author{Ying Li}
\affiliation{Department of Materials, University of Oxford, Parks Road, Oxford OX1 3PH, United Kingdom}

\author{Andrew M. Steane}
\affiliation{Clarendon Laboratory, Department of Physics, University of Oxford, Parks Road, Oxford OX1 3PU, United Kingdom}

\author{Daniel Bedingham}
\affiliation{Faculty of Philosophy, University of Oxford, Woodstock Road, OX2 6GG, United Kingdom.}

\author{G. Andrew D. Briggs}
\affiliation{Department of Materials, University of Oxford, Parks Road, Oxford OX1 3PH, United Kingdom}

\date{\today}

\begin{abstract}
Continuous spontaneous localisation (CSL) is a model that captures the effects of a class of extensions to quantum theory which are expected to result from quantum gravity, and is such that wavefunction collapse is a physical process. The rate of such a process could be very much lower than the upper bounds set by searches to date, and yet still modify greatly the interpretation of quantum mechanics and solve the quantum measurement problem. Consequently experiments are sought to explore this. We describe an experiment that has the potential to extend sensitivity to CSL by many orders of magnitude. The method is to detect heating of the motion of charged macroscopic objects confined in a Paul trap. We discuss the detection and the chief noise sources. We find that CSL with standard parameters could be observed using a vibration-isolated ion trap of size $1\,{\rm cm}$ at ultra-low pressure, with optical interferometric detection.
\end{abstract}

\maketitle

Since quantum technologies have developed rapidly in recent years, it is becoming possible to test quantum theory on macroscopic objects. One of the interesting questions that arises is that of wavefunction collapse. While superposition states of microscopic objects have been observed in many experiments, and such tests have been extended to mesoscopic systems such as large molecules~\cite{Eib2013}, observations on macroscopic objects have, to date, been ambiguous, since the lack of observable interference could be attributed either to a fundamental collapse effect or to a lack of experimental precision. However, the possibility of a true collapse process in nature is attractive because such a process is expected to arise naturally in models of some sort of quantised nature of spacetime~\cite{Penrose1996}, and because it would provide a solution to the quantum measurement problem. The latter is the problem famously illustrated by Schr\"odinger's cat, namely that it is hard to reconcile our everyday experience with the implications of quantum theory as currently understood.

One of the theoretical efforts to account for wavefunction collapse is the continuous spontaneous localisation (CSL) model~\cite{Ghirardi1986,Ghirardi1990,Ghirardi1995,Bassi2003,Bassi2013,Collett2003}. In this model wavefunction collapse occurs spontaneously, and the collapse strength is proportional to mass. It thus predicts that long-lived spatial superposition states of massive objects would not be found in nature.

Potential phenomena due to CSL include the decoherence of a superposition state~\cite{Romero2011PRL,Romero2011,Pepper2012}, linewidth broadening~\cite{Bahrami2014} and heating of a mechanical oscillator~\cite{Bassi2005,Adler2004,Adler2005,Vinante2016}, and diffusion in free space~\cite{Collett2003,Bera2014}. All these can also be caused by ordinary decoherence associated with uncontrolled interactions with the environment, therefore low noise is the crucial consideration in the design of any experiment in this area. In this paper we investigate the potential of a Paul trap (often called `ion trap') to provide the low-noise environment that is required. We develop an idea of Collett and Pearle's, in which one seeks to detect the heating due to CSL of a charged macroscopic object confined in such a trap~\cite{Collett2003,Millen2015}. We evaluate the noise sources and detection possibilities sufficiently well to demonstrate the feasibility of the method. We show that CSL with parameters suggested by Ghiradi, Rimini and Weber (GRW)~\cite{Ghirardi1986} could be detected in a trap of size $\sim 1 \,{\rm cm}$ at a pressure $10^{-13}  \,{\rm Pa}$ in a time of order one minute. 

\begin{figure}[tbp]
\centering
\includegraphics[width=1\linewidth]{\figpath 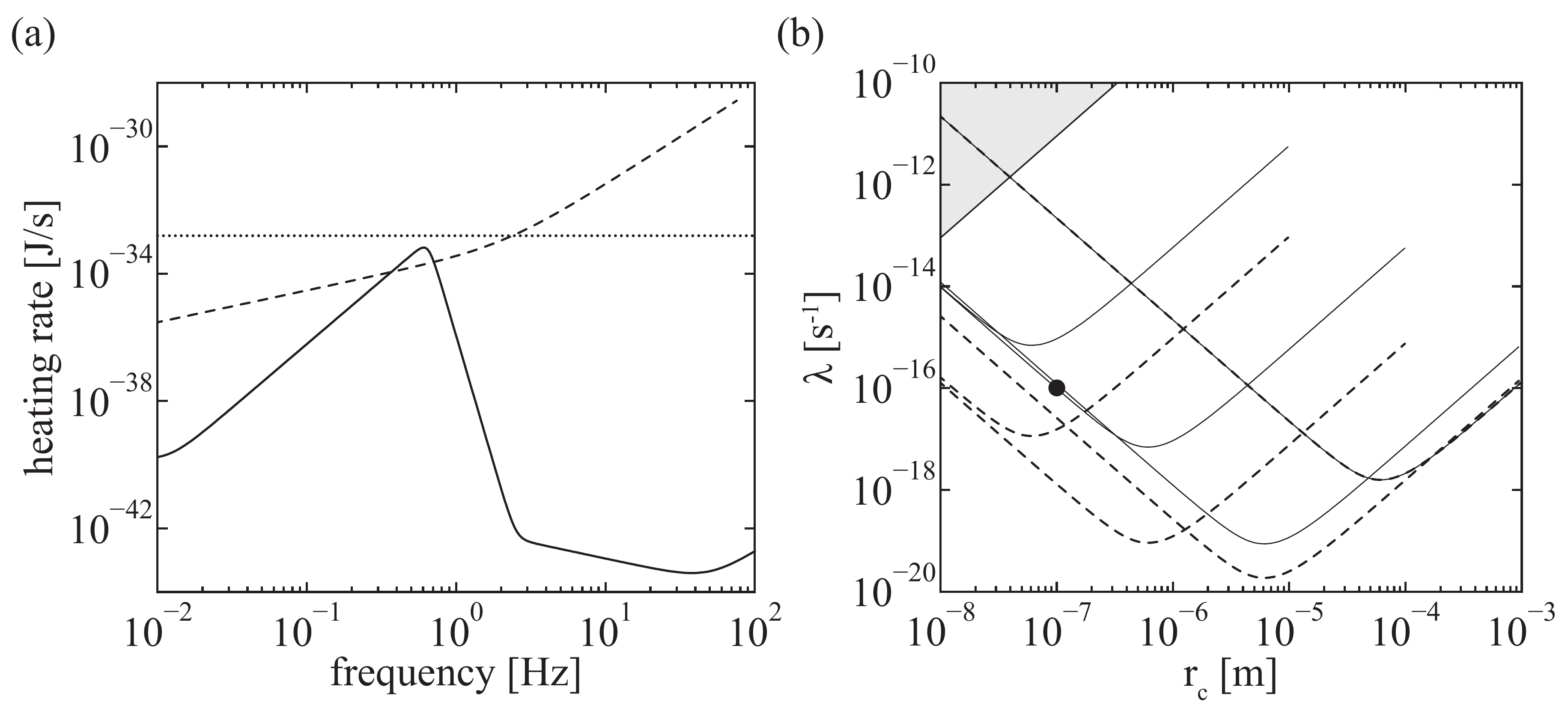}
\caption{
Heating rate and detection limit. (a) Heating rate owing to various processes, calculated for a singly-charged osmium sphere of radius $0.238 \,\mu{\rm m}$, mass $7.7 \times 10^{11}$ atomic units. Full line: $\Gamma_x$ (mechanical), dashes: $\Gamma_E$ (electric field), dots: CSL heating $\Upsilon$ at GRW parameters. (b) Smallest detectable value of $\lambda$, as a function of $r_{\rm c}$, for four sizes of a trapped osmium sphere: $L=0.1,1,10,100 \,\mu{\rm m}$. The minimum of the corresponding curve is in the vicinity of $L$. The continuous curves are for background pressure $p=10^{-12} \,{\rm Pa}$, the dashed curves for $p=10^{-14} \,{\rm Pa}$. Two curves for $L = 100 \,\mu{\rm m}$ are overlapped. The shaded region is the currently excluded region, set by spontaneous x-ray emission~\cite{Curceanu2015}. The dot indicates GRW values.
}
\label{fig:noise}
\end{figure}

CSL is characterised by two parameters: the collapse rate of a nucleon $\lambda$ and the critical length scale $r_{\rm c}$~\cite{Collett2003}. An equivalent set of parameters $\gamma = (4\pi r_{\rm c}^2)^{3/2}\lambda$ and $\alpha = r_{\rm c}^{-2}$ may also be used~\cite{Bassi2013}. A standard choice is $\lambda \sim 10^{-16}  \,{\rm s}^{-1}$ and $r_{\rm c} \sim 10^{-7}  \,{\rm m}$; we will refer to these as `GRW values'~\cite{Ghirardi1986}. The range of values not yet excluded by observations extends very much higher than this [c.f. Fig.~\ref{fig:noise}(b)]~\cite{Bassi2013,Vinante2016}, so that even to approach a sensitivity sufficient to detect GRW values would be a significant achievement.

\begin{figure}[tbp]
\centering
\includegraphics[width=1\linewidth]{\figpath 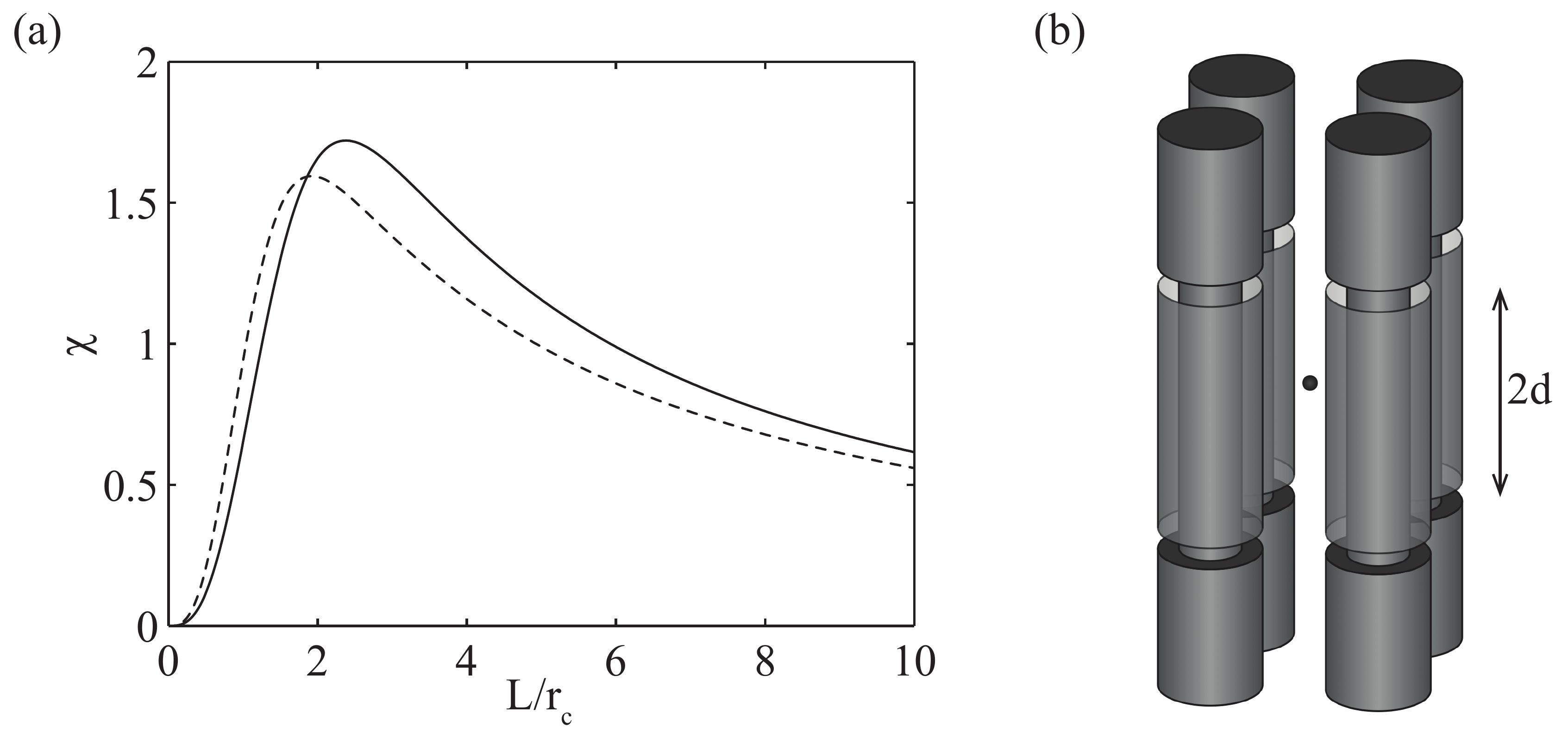}
\caption{
Energy raising rate and Paul trap. (a) CSL factor $\chi$ of a sphere of radius $L$ in free space (full line), and a cube of side $2L$ in a 1D harmonic potential (dashed). (b) Example electrode structure for a Paul trap.
}
\label{fig:chi}
\end{figure}

One effect due to CSL of a rigid body is to raise the energy in the centre of mass motional degrees of freedom (CMM). The energy raising rate (ERR) (roughly speaking, the heating rate) can be written in the form
\be
\Upsilon = \chi {\hbar^2 \lambda r_{\rm c} \rho}{u^{-2}},
\ee
where $\rho$ is the density of the material, $u$ is the mass of a nucleon, and the dimensionless factor $\chi$ depends on the shape of the rigid body and the external potential. For a sphere of radius $L$ in free space, $\chi = 2\pi I/x$ for each of the three motional directions, where $x \equiv L/r_{\rm c}$, and~\cite{Collett2003}
\be
I = 1 - 2x^{-2} + \left( 1 + 2x^{-2} \right) \exp(-x^2).
\ee
The maximum value is $\chi \simeq 1.7202$ at $L \simeq 2.38 r_{\rm c}$ [see Fig.~\ref{fig:chi}(a)]. For a cube of side $2L$ in a one-dimensional (1D) harmonic trap, $\chi = {I_{12} I_3}/x^3$~\cite{Bassi2005,Adler2004,Adler2005}, where $I_3  = 2[ 1 - \exp(-x^2) ]$ and
\be 
I_{12} = \left[ \exp(-x^2) - 1  +  \sqrt{\pi} \, x{\rm Erf}(x) \right]^2.
\ee
Here, ${\rm Erf}(x) \equiv 2 \pi^{-1/2}\int_0^x \exp(-t^2) {\rm d}t$ is the error function. For the cube, $\chi$ is maximised at $L \simeq 1.92 r_{\rm c}$ with the value $\chi \simeq 1.5943$.

Taking GRW parameter values and $\rho = 22,587 \,{\rm kg}/{\rm m}^3$ (the density of osmium), the maximum ERR for a sphere in free space is $\Upsilon \simeq 1.57 \times 10^{-33} \,{\rm J}/{\rm s}$, or $6.8$ nanokelvin per minute and for a cube it is a little lower. The essence of the proposed experiment is to trap and cool an object of this size, and then determine whether its CMM temperature in 1D has increased by $10 \,{\rm nK}$ after $90$ seconds (or $100 \,{\rm nK}$ after 15 minutes)~\cite{diskcomment}.

There are two main experimental issues. First, can one construct an apparatus in which the heating caused by other noise sources does not dominate that due to CSL? Secondly, can one detect a heating rate of this order in a reasonably short amount of time?

The main contributions to heating are from mechanical vibrations, electric field noise, magnetic field noise and background gas collisions. Electric dipole radiation from the oscillating charged rigid body is negligible, as is the momentum diffusion caused by scattering black body radiation at $70$ kelvin~\cite{Chang2010}.

Suppose a change $\delta R$ in a stochastic variable $R$ causes the force on the trapped body to vary by $\delta F$. Then the fluctuations of $R$ cause an average heating rate given by
\be
\Gamma_R = \frac{1}{4 m} \left( \pd{F}{R} \right)^2 S_R(\omega_0),
\label{GamR}
\ee
where $m$ is the mass of the trapped body, $\omega_0$ is the angular frequency of its simple harmonic motion, and $S_R$ is the one-sided power spectrum of the noise in $R$.

First consider mechanical noise, in which the trap electrodes are displaced by a distance $x$, for example owing to seismic noise and thermal vibrations of the electrode surfaces. Then $\partial F/\partial x = m \omega_0^2$, so the heating rate owing to these fluctuations is $\Gamma_x = m \omega_0^4 S_x/4$, where $S_x$ is the positional noise. The trap electrodes (to be discussed) will have dimensions in the range millimetres to metres, and we require stability of the apparatus relative to a local inertial frame (that is, the apparatus should maintain, as far as possible, a fixed acceleration relative to a frame falling freely in the local gravitational field). There are two main techniques to achieve high stability: active methods based on laser interferometry, and passive methods based on mechanical filters, which are low-frequency resonators (e.g.~pendulums or masses on springs). Above its resonance frequency $f_0$ the transmission of each filter is proportional to $(f_0/f)^{2}$, and by cascading filters one obtains higher powers. It will emerge that the frequency scale we are interested in here is in the vicinity of that used in gravitational wave sensors such as LIGO and VIRGO, and these illustrate the state of the art. In an ordinary optics lab, optical tables can be stabilised to a level of $10^{-10} \,{\rm m}/\sqrt{{\rm Hz}}$ between $10 \,{\rm mHz}$ and $100 \,{\rm Hz}$~\cite{Dahl2012}, or somewhat better. The advanced LIGO positioning platform is designed to achieve $10^{-11} \,{\rm m}/\sqrt{{\rm Hz}}$ at $1 \,{\rm Hz}$ by active stabilisation~\cite{Matichard2015}. The mechanical filters then achieve lower noise at higher frequencies. For the purpose of this study we shall assume the noise is about an order of magnitude larger than that measured at the LIGO interferometers~\cite{LIGOpaper}, and is given by the following model, which reproduces the design study for the VIRGO experiment described in~\cite{Blom2015}:
\be
S_x = a_1^2 f^{-5} + a_2^2 (f_0^{20}+ f^{20})^{-1} + a_3^2 f^{-1}
\ee
where $f_0=0.65 \,{\rm Hz}$ and if $f$ is in ${\rm Hz}$ and $S_x$ is in units of ${\rm m}^2/{\rm Hz}$ then the coefficients are given by $a_i = \{1.5,\,1500,\,0.0006\} \times 10^{-15}$. The three terms model the effects of thermal vibrations in the pendulum suspensions, seismic vibrations after filtering, and thermal vibrations of the electrode surfaces, respectively. Below $f_0$ this gives $S_x \simeq 10^{-20} \,{\rm m}^2/{\rm Hz}$, achieved by active stabilisation. The resulting heating rate for an example mass in a harmonic trap is shown, as a function of frequency, by the full line in Fig.~\ref{fig:noise}(a).

The conclusion so far is that mechanical noise can be suppressed sufficiently for it to produce a heating rate small compared to the CSL effect ($\Upsilon$) at the GRW parameter values, and we have a described a model which allows us to explore a range of parameter values.

Next, consider electric field noise. The force owing to an electric field fluctuation $\delta E$ is $q \delta E$ where $q$ is the charge of the trapped object, so the heating rate is $\Gamma_E = (q^2/4m) S_E(\omega_0)$. This neglects cross-coupling between noise fields and the rf trapping fields, which is acceptable since the latter makes a small correction and we only need an approximate estimate. The electric field noise in a Paul trap is notoriously difficult to calculate from models of the materials and electronic sources which drive the trap, but it can be estimated with reasonable confidence by using existing data from a number of experiments. Typically one measures the heating rate of a trapped ion, and infers $S_E$. This has been done for a large number of traps with sizes between a fraction of a millimetre and a centimetre, in experiments designed for low noise. The vibrational frequency of the trapped ion is typically in the range $10^5$--$10^6 \,{\rm Hz}$ in such experiments, so they do not explore the low frequency noise which we are interested in. The observations are consistent with a noise $S_E$ scaling as $\omega^{-n}$ with $0 < n < 1$, for the rather restricted range of frequencies typically studied in any given trap. We will assume the scaling is given by $n=1$, and adopt the following model:
\be
S_E =  \left[ (b_1 + b_2 V_Q^2)d^{-2} + b_3 d^{-4} \right] {\omega}^{-1}
\label{SE}
\ee
where $b_i$ are coefficients, $V_Q$ is the applied voltage that produces the constant quadrupole field, and $d$ is the distance from one of the electrode surfaces to the centre of the trap [Fig.~\ref{fig:chi}(b)]. The terms $b_1$ and $b_2$ allow for contributions to the noise both unrelated to $V_Q$ and increasing with $V_Q$. The terms in $d^{-2}$ describe noise owing to fluctuation of the voltage on the whole surface of any one electrode, the term in $d^{-4}$ describes noise owing to voltage fluctuations in a collection of independently fluctuating patches, where each patch has a size small compared to the distance to the point where the field is measured.

The value of the coefficient $b_3$ is strongly dependent on the temperature of the electrode surfaces~\cite{Lab2008}, and one would expect some temperature dependence in $b_1$ also. In the present study we will assume the electrodes are at room temperature, and we will typically be interested in `large' traps where the $b_3$ term is negligible (though we include it anyway). The data reported in~\cite{Poulsen2012} are approximately reproduced by the parameter values $b_1 = 1.7\times 10^{-14} \,{\rm V}^2$, $b_2 = 1.1 \times 10^{-17}$, $b_3 = 2.6 \times 10^{-19} \,{\rm V}^2{\rm m}^2$. To arrive at these values, we estimated the relative contributions of the $b_1$ and $b_2$ terms; these will vary from case to case but our deductions give a sufficient estimate of $\Gamma_E$ to determine whether the proposed experiment is feasible.

In thus using the data from existing experiments, we will be making a large extrapolation in frequency (from $10^5 \,{\rm Hz}$ to $0.1 \,{\rm Hz}$), but since we assume $1/f$ noise this is a conservative estimate. The model predicts that the field noise below $1 \,{\rm Hz}$ is similar to that which would be given by voltage noise of order some tens of nanovolts for applied voltages of order tens of volts; this is a reasonable value, since we are only concerned with the noise component remaining after common mode rejection, and each pair of end cap electrodes can be made of a single lump of metal~\cite{Boggs1995}. The r.m.s. electric field we thus obtain is similar to the electric field that would be produced at the trap centre by a single electron on an electrode surface.

The dashed curve in Fig.~\ref{fig:noise}(a) shows the heating rate owing to electric field noise for an example case. Here we specify values of $d$ in the range $0.1 \,{\rm mm}$ to $1\,{\rm m}$, and adopt $V_Q = 20$ volt. We then calculate the vibrational frequency $\omega_0$, which scales as $1/d$, and obtain $S_E(\omega_0)$ from Eq.~(\ref{SE}). The result of the scaling with $d$ is that lower frequencies correspond to larger traps and consequently less noise. We thus deduce that, for vibrational frequencies below about $1 \,{\rm Hz}$, which here corresponds to a trap of size $d \simeq 1 \,{\rm cm}$, the electric field noise produces less heating than CSL.

Magnetic field noise can heat the CMM by coupling to the current associated with the oscillating charge, or to the magnetic dipole moment $\mu$ of the rigid body itself. The former contribution is negligible; the latter affects neutral or charged bodies equally, as follows. A fluctuation in the $B$-field {\em gradient} is conservatively estimated as $(\delta B/d)$; the associated CMM heating is given by Eq.~(\ref{GamR}) using $\partial F/\partial B \approx \mu / d$. After modest precautions, and without superconducting shields, the r.m.s. magnetic field noise observed in ion trap experiments is of order $100 \,\mu{\rm G}$ (i.e.~$10^{-8}$ tesla) for a bandwidth of order $1 \,{\rm kHz}$, which suggests $S_B \approx 10^{-19} \,{\rm T}^2/{\rm Hz}$. Using this value and the sphere as described, we find this effect is negligible for $\mu \ll 10^7$ Bohr magnetons for $d = 1 \,{\rm cm}$; expected values of $\mu$ are comfortably in this region.

If the trapped body is conducting, the a.c. field of the Paul trap creates currents in it, and these can couple to magnetic field noise. The induced electric dipole between one part of the sphere and another is of order $d_0 \simeq \epsilon_0 L^3 E$ where $E \simeq Q_{\rm AC} L$ is the electric field at the sphere's surface owing to the trap quadrupole $Q_{\rm AC} \simeq V_{\rm AC}/d^2$. The magnetic force on the induced currents is approximately $\Omega_{\rm AC} d_0 B$ where $\Omega_{\rm AC}$ is the Paul trap drive frequency. In the absence of asymmetry, these forces on different parts of the sphere balance, but even if this were not so the resulting heating rate would be of order $(\Omega_{\rm AC} d_0)^2 S_{\rm B}/4m \approx 10^{-65} \,{\rm J}/{\rm s}$, which is negligible.

Next we consider the effect of collisions with background molecules in the vacuum chamber. The effect depends on whether individual collisions can be detected. If they cannot, then they provide diffusive heating at the rate $\Gamma_c = (m_{\rm g}/m) p \sigma \bar{v}$, where $m_{\rm g}$ is the mass of a background atom or molecule, $p$ is the pressure, and $\sigma = 2\pi L^2$ is the collision cross-section for heating in 1D of a sphere of radius $L$. We thus find an upper bound on $p$ in order that collisional heating should be smaller than $\Upsilon$:
\be
p < \Upsilon m / \left(2 \pi L^2 \bar{v} m_{\rm g}\right).
\ee 
At the low pressure required, cryogenic pumping is needed, and therefore the residual gas is mostly light gases such as hydrogen and helium. Taking helium at room temperature, and GRW parameters, we find $p < 7 \times 10^{-13} \,{\rm Pa}$. This is challenging but possible (the lowest reported pressure is around $7 \times 10^{-15} \,{\rm Pa}$~\cite{Collett2003,Gabrielse1990}).

At a pressure of $10^{-13} \,{\rm Pa}$, the collision rate is of order one per 90 seconds. 

So far we have shown that the expected noise sources do not dominate CSL in the significant parameter regime. It remains to explore whether or not the CSL effect is itself measurable. That is, is it feasible to detect some tens to hundreds of nanokelvin of heating of the motion of the trapped rigid body? One can readily suggest experimental methods to show that it is. One may detect the position of the body in 1D by reflecting a laser beam off it and using an interferometric method. By repeating the measurement after a quarter cycle of the oscillation in the trap, one locates the body in phase space. A natural limit of such methods is the `standard quantum limit'~\cite{Caves1980}. By studying phase estimation from scattered light pulses, one finds that in principle the 1D position and momentum of a measured object can be determined to within an area of order $\hbar$ in phase space. Such a precision, when combined with feedback to the trap electrodes, amounts to the ability to prepare the object near to its ground state of motion, which is in practice very difficult to achieve for macroscopic objects. However we do not need to assume that precision here. Suppose that in any given run of the experiment one achieves measurement precisions $\Delta x$ and  $\Delta p = m \omega_0 \Delta x$, such that $\Delta x \Delta p = 2 \bar{n} \hbar$, for some $\bar{n} \gg 1$. Then the initial state of the sphere can be prepared with energy $E_0 = (1/2) m \omega_0^2 \Delta x^2 = (1/2) \omega_0 \Delta p \Delta x = \bar{n}\hbar \omega_0$, and one can also detect energy increases of this order. At $\bar{n} = 500$, for example, $E_0 / \kB = 2.4 \,{\rm nK}$ when $\omega_0/2\pi = 0.1 \,{\rm Hz}$. CSL will give this amount of heating in 23 seconds.

So far we have surveyed the potential to detect CSL at the parameters suggested by GRW. In practice one would test rigid bodies of a number of different sizes in order to explore different regions of parameter space. One would also accumulate data over long runs. Our study has shown that even a single run of duration of order one minute could detect CSL at the GRW parameters. A data set accumulated over longer periods would provide the means to check for systematic errors, as well as to reduce the statistical uncertainty.

By using the noise models discussed, we can map the region of parameter space in which CSL could be detetected by this approach, see Fig.~\ref{fig:noise}(b). The boundaries indicated are only approximate, but they suffice to show that this is a feasible experimental technique, which could set an upper bound on $\lambda$ many orders of magnitude below the currently known upper bound, or indeed discover CSL. The concept of the method is due to others~\cite{Collett2003}; the contribution of the present work is to discover suitable parameters for the apparatus and to show that the concept is sound. The results shown in Fig.~\ref{fig:noise}(b) are mostly limited by collisions, and therefore are not very sensitive to the trap electrode size and voltage. A size $d \sim 1-10 \,{\rm cm}$ is suitable for $L < 10 \,\mu{\rm m}$; for larger spheres a smaller trap is useful in order to keep the vibrational frequency above $0.01 \,{\rm Hz}$ (below this we would hesitate to trust our noise models).

We have discussed effects which increase the kinetic energy of the collapsing object. An apparatus based on the same Paul trap would be a good starting point for more sophisticated experiments designed to detect motional decoherence. Superpositions of motional states could be produced by adapting various techniques that have been applied to single trapped ions~\cite{Wineland2013}. For example, one could use a diamond sphere with a nitrogen-vacancy center, or a semiconducting sphere containing a quantum dot, and obtain a spin-dependent dipole force from the interaction with an optical standing wave. More generally, one would seek a controlled coupling between the centre of mass of the trapped microsphere and some other entity or degree of freedom which can be prepared in a superposition. The CSL family of phenomenological theories offers the potential to solve what many recognize as the measurement problem, not so much by a modification to quantum mechanics, as a way of capturing what might naturally arise in various possible frameworks of quantum theories with dynamic spacetime~\cite{Briggs2013}. It may soon prove possible to change the bounds of the two key parameters using data from practical laboratory scale experiments.

\begin{acknowledgments}
We thank Tom Harty for helpful information. This project was made possible through grants from EPSRC (EP/J015067/1) and Templeton World Charity Foundation. The opinions expressed in this publication are those of the authors and do not necessarily reflect the views of Templeton World Charity Foundation.
\end{acknowledgments}


\begin{thebibliography}{9}
\bibitem{diskcomment} We also considered rotational diffusion of a disk, but abandoned this, owing to problems associated with the oscillating torque from the trapping fields.
\end{thebibliography}
\end{document}